# Spectroscopic observations of the bright RV Tauri variable R Scuti

## David Boyd

A series of spectra of the RV Tauri star R Scuti taken as it rose from a deep minimum to a bright maximum during 2013 October and November reveals major changes taking place in the photosphere and outer atmosphere of the star. This may be the first such series of spectra and demonstrates the capability of amateur spectroscopy for studying these complex stars.

## Introduction

RV Tauri stars are radially-pulsating yellow supergiant stars[1] and are located near the top of the instability strip on the Hertzsprung–Russell diagram. The RV Tauri stage is a relatively short-lived episode late in the lifetime of these stars as they transition from the asymptotic giant branch (AGB) towards eventually becoming a white dwarf. Their lightcurves are characterised by alternating deep and shallow minima, although this pattern is not always strictly followed.

R Scuti is one of the brightest RV Tauri variables. Its visual magnitude ranges from around 4.7 at maximum to 6.0 during a shallow minimum and below 8.0 at a deep minimum.[2] On 2013 October 1 John Toone reported on the BAAVSS-alert email list that R Sct was undergoing a deep minimum immediately following the longest primary minimum on record. He has since described his visual observations during 2013 in a BAAVSS *Circular* article.[3]

I thought it would be interesting to follow the star's rise from minimum spectroscopically. Further investigation showed that R Sct has an intriguing spectroscopic history.

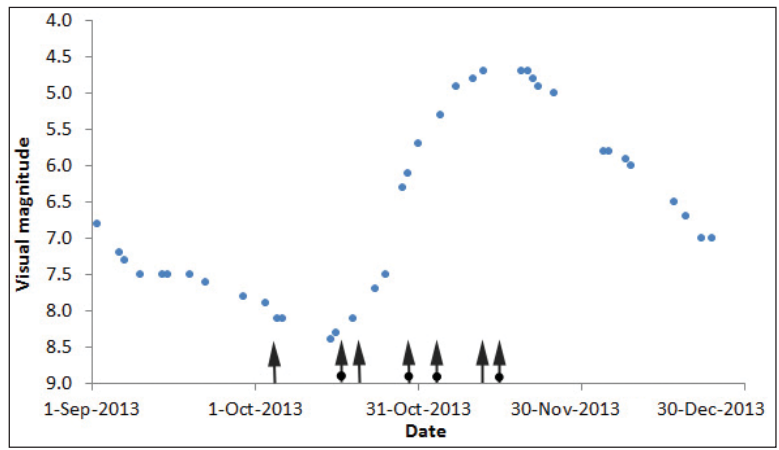

**Figure 1.** Visual lightcurve of R Sct recorded by Toone[2] with arrows indicating when spectra were recorded. The dotted arrows indicate the spectra shown in Figures 2–5.

## The mysterious spectrum of R Sct

Visual observations of the spectrum of R Sct were reported by T. E. Espin in *Monthly Notices* of the RAS in 1890.[4] Espin commented that the bright lines he could see in the spectrum changed as the star varied in brightness. The first photographic spectroscopy of R Sct was carried out by Curtiss at Lick Observatory in 1903.[5] He noted the short-lived appearance of bright hydrogen emission lines. In 1922 Joy[6] reported that the spectrum varied between spectral types G5 and K2, but that titanium oxide (TiO) molecular absorption bands appeared when the star was fainter than about mag 5.7. This was unexpected, as TiO bands are normally only seen in stars of spectral type M.

Spectroscopic observations were continued in the 1930s at the University of Michigan Observatory by McLaughlin using a single prism spectrograph attached to a 37-inch [94 cm] reflector.[7] He confirmed the findings of Joy and noted that the deeper the minimum, the stronger the TiO bands, with the spectrum becoming more reminiscent of an M-type star at a deep minimum. He confirmed earlier reports that hydrogen emission lines appeared as the star brightened from minimum, then faded rapidly around maximum light. He interpreted the changing brightness and spectrum of the star in terms of physical pulsations, with the temperature and brightness of the star increasing as it contracted and falling as it expanded.

In 1984, Wallerstein & Cox[8] attributed the presence of hydrogen emission lines in pulsating stars to the passage of a shock-wave. This was later confirmed in high-resolution spectroscopic observations of R Sct by Lèbre & Gillet in 1991[9] who showed that the shock-wave corresponded to a photospheric acceleration peak prior to minimum luminosity which gave rise to hydrogen-alpha emission as the star subsequently brightened.

On 1981 September 26, Howell *et al*.[10] fortuitously recorded a high-dispersion spectrum of R Sct when it was in a deep minimum at around 7.9mv. In 1982 July they managed to get further spectra with the star at a similar magnitude. These spectra were dominated by strong TiO absorption bands.

In a paper in the AAVSO *Journal* in 1985, Wing[11] attributed the apparent split spectral personality of R Sct, in which it appeared to have the characteristics of both K and M spectral types at the same time, as being due to a greatly extended atmosphere with cool upper layers loosely attached to the photosphere. While the photosphere produces a spectrum which varies between types G and K as the star pulsates, the upper atmosphere is about 1500K cooler. When the star is at minimum brightness, and therefore at its coolest, TiO molecules can form in the upper atmosphere and introduce their absorption bands into the photospheric spectrum. The deeper the minimum, the more prominent the TiO bands until, in a very deep minimum, they dominate the spectrum. Observations of the infrared spectrum of R Sct reported by Hinkle *et al.*[12] provided insight into the complex pulsational motions and physical conditions at various depths in the atmosphere.

**Table 1.** Times of spectra of R Sct and its visual magnitude at or near the same time.

| Times of spectra | Visual magnitude |
|---|---|
| 2013 Oct  5.775 | 8.1 |
| Oct 17.764 | 8.2 |
| Oct 20.772 | 8.0 |
| Oct 29.781 | 6.1 |
| Nov  4.727 | 5.3 |
| Nov 12.748 | 4.7 |
| Nov 14.727 | 4.7 |

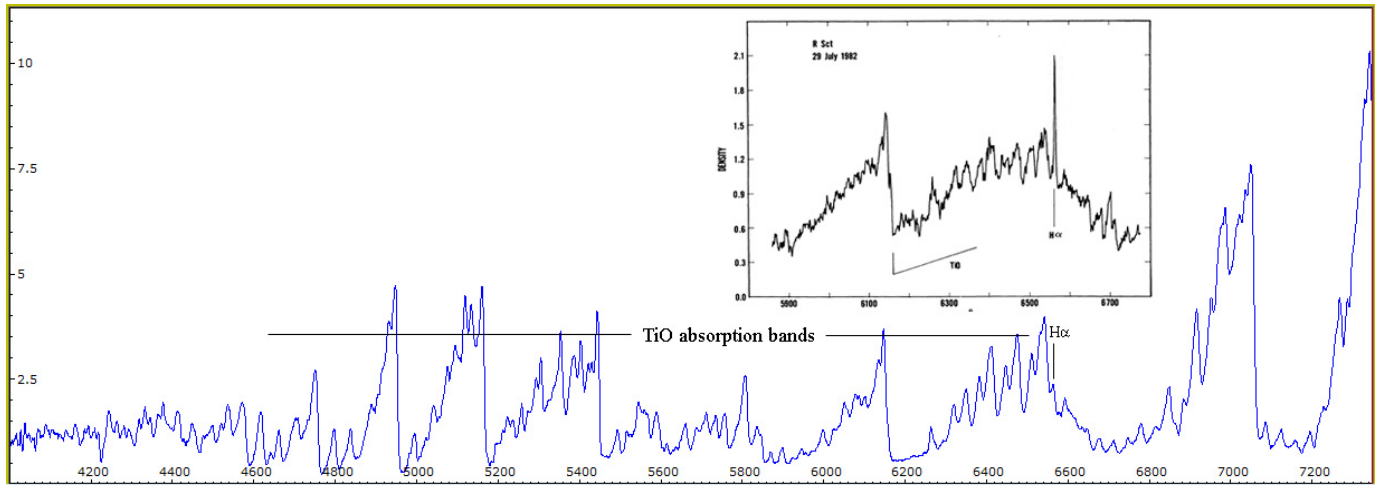

**Figure 2.** Spectrum of R Sct on 2013 Oct 17.764 at minimum light with strong TiO molecular absorption bands and (inset) spectrum from Howell *et al.*[10] 14 days past minimum.

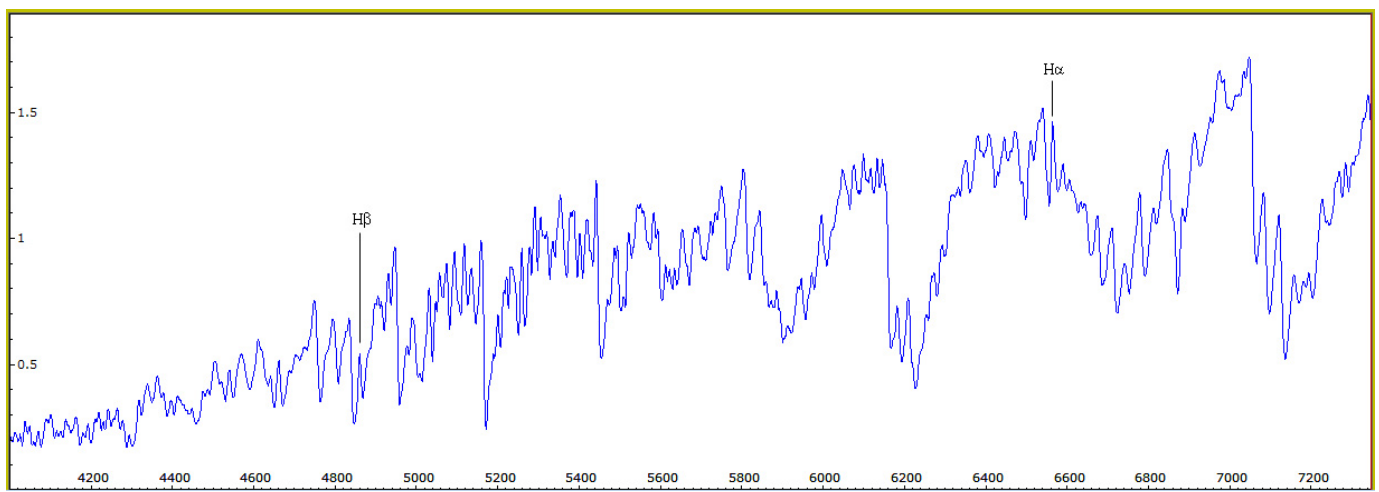

**Figure 3.** Spectrum of R Sct on 2013 Oct 29.781 as the star increased in brightness, the TiO bands became weaker and the H-alpha line started to increase.

In summary, the behaviour of R Sct is still not fully understood although its interpretation as a large, radially-pulsating, post-AGB star appears to be generally accepted. Understanding the composite nature of the spectrum of R Sct is still a work in progress. One paper describes R Sct as being '*in a state of deterministic chaos*'.

## Spectroscopic observations of the deep minimum of 2013 October

Following Toone's email alert on 2013 Oct 1, I obtained spectra of the star on 7 nights between Oct 5 and Nov 14 using a LISA spectrograph attached to a Celestron 280mm SCT. The spectra have a resolution of 7Å and a signal-to-noise ratio in excess of 700 at the H-alpha wavelength. All spectra were wavelength calibrated with the internal neon lamp plus Balmer absorption lines and corrected for instrumental and atmospheric extinction effects using a nearby A-type star.

The dates of the spectra are listed in Table 1 together with the visual magnitude reported by Toone at or near the same time as recorded in the BAAVSS database.[2] Figure 1 shows Toone's visual lightcurve with arrows indicating when spectra were recorded. During this period R Sct faded to a minimum of 8.4mv, then rose in 28 days to 4.7mv.

Four of these spectra marked with dotted arrows in Figure 1 are shown in Figures 2–5. The horizontal axes give the wavelength in Ångstroms, and the vertical axes represent intensity with each spectrum normalised to unity at 6650Å.

The spectrum in Figure 2 was recorded on 2013 Oct 17 at the bottom of the deep minimum. By comparison with spectra from the Pickles *Stellar Spectral Flux Library*,[13] the general characteristics of the spectrum are similar to a mid-M spectral type. Also shown (inset) is the spectrum taken by Howell *et al.* on 1982 July 29 (Figure 5 from ref.10) about 14 days past primary minimum. Both spectra show strong TiO molecular absorption bands. Howell's spectrum also shows a prominent H-alpha emission line which is much less prominent in my spectrum. The spectra I recorded on Oct 5 and Oct 20 also show strong TiO absorption bands and a similarly weak H-alpha emission line.

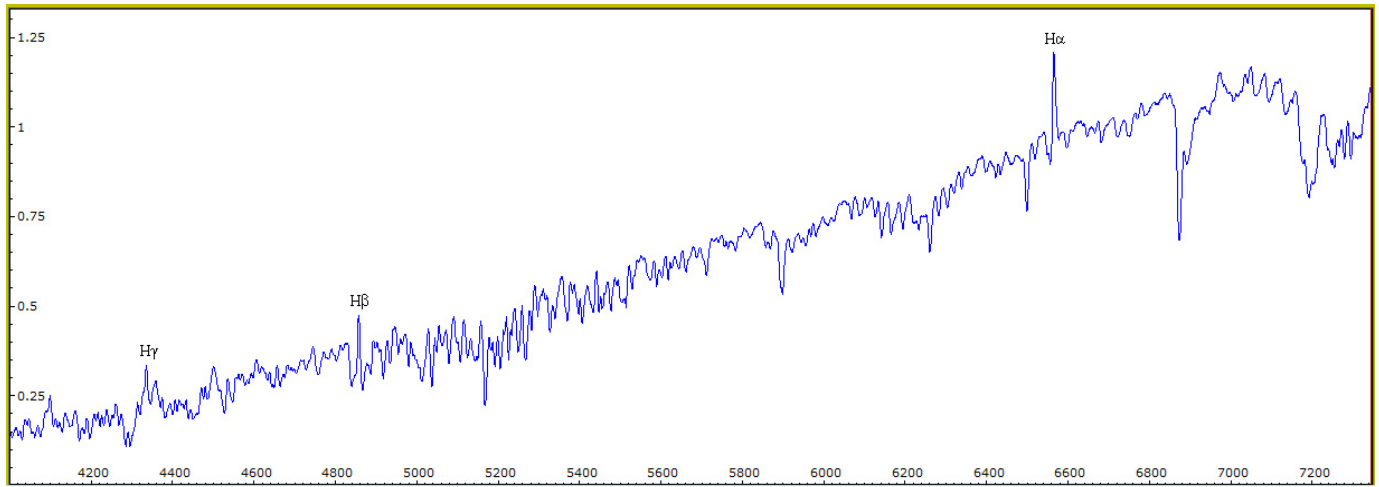

**Figure 4.**  Spectrum of R Sct on 2013 Nov 4.727 with the hydrogen Balmer series lines now clearly visible in emission.

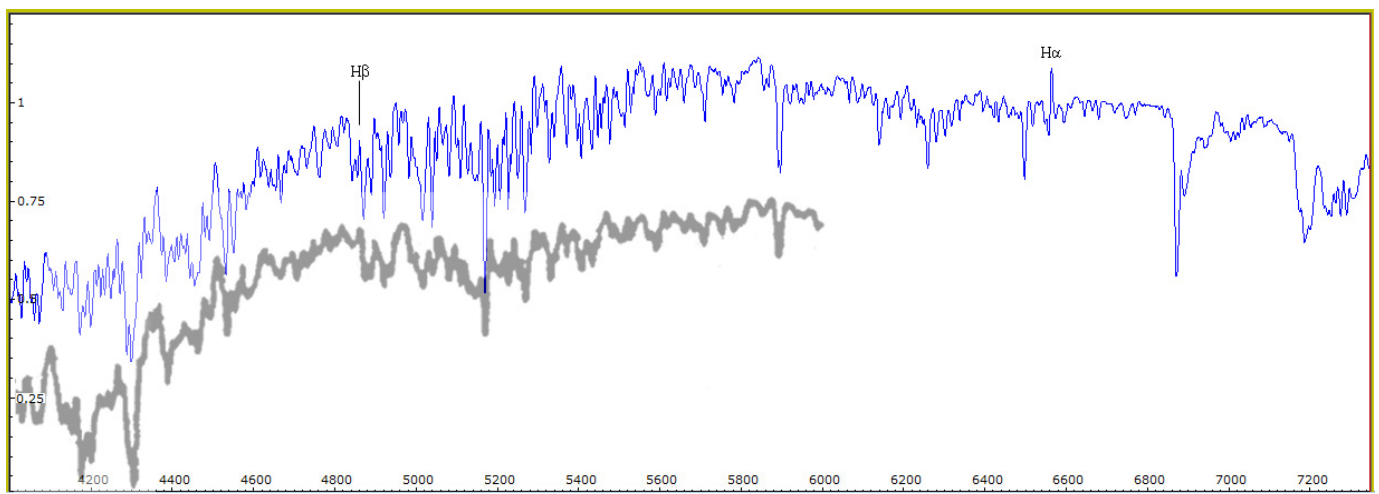

**Figure 5.**  Spectrum of R Sct on 2013 Nov 14.727 at maximum light, with the corresponding segment of a spectrum from Shenton *et al.*[14] when R Sct was in a similarly bright state. The latter is displaced vertically for clarity.

Twelve days later on Oct 29 (Figure 3) the TiO bands are weaker and the H-alpha line is stronger. This spectrum is now closer to early M-type.

A further 5 days later on Nov 4 (Figure 4), as the star rapidly brightens, the TiO bands have almost disappeared and the spectrum shows hydrogen Balmer series emission lines as reported by several observers in the past. The spectrum is now a complex mixture of mid-K and early M spectral types. Finally, on Nov 14 (Figure 5), the star has reached maximum brightness and the spectral type is now a combination of mid/late G and early K. Also shown for comparison in Figure 5 is a segment of a low resolution spectrum recorded by Shenton *et al.* (Figure 8(a) from ref.14) when the star was in a bright state. The two spectra are very similar.

When all four spectra are taken together, the progressive shift in the energy balance of the spectrum continuum towards shorter wavelengths as the star brightens and its temperature increases is clear. This is probably the only set of visual spectra of R Sct obtained as it rises from a very deep minimum to a bright maximum. It is noteworthy that the spectroscopic behaviour of these complex stars, which has in the past been the exclusive province of professional observatories, can now be studied with backyard amateur equipment.

## Acknowledgment

I thank the two referees, Dr Chris Lloyd and Robin Leadbeater, for their helpful comments which have improved the paper.

**Address:**  5 Silver Lane, West Challow, Wantage, Oxon. OX12 9TX. [drsboyd@dsl.pipex.com]